\documentclass[
showpacs,preprintnumbers,amsmath,amssymb,11pt]{revtex4}
\usepackage{epsfig,graphicx}

\def\wt{\widetilde}

\begin{document}

\tolerance=5000

\title{Renormalization-group inflationary scalar electrodynamics
and $SU(5)$ scenarios confronted with Planck2013 and BICEP2 results
}\author{E.~Elizalde$^{1}$\footnote{E-mail address:
elizalde@ieec.uab.es}, \
S.D.~Odintsov$^{1,2,3}$\footnote{E-mail address:
odintsov@ieec.uab.es}, \
E.O.~Pozdeeva$^{4}$\footnote{E-mail address:
pozdeeva@www-hep.sinp.msu.ru}, \
S.Yu.~Vernov$^{4}$\footnote{E-mail address:
svernov@theory.sinp.msu.ru}}
\affiliation{$^1$Instituto de Ciencias del Espacio (ICE/CSIC) \, and
  Institut d'Estudis Espacials de Catalunya (IEEC)
Campus UAB, Facultat de Ci\`encies, Torre C5-Parell-2a
planta,
 08193 Bellaterra (Barcelona), Spain\\
$^2$Instituci\'o Catalana de Recerca i Estudis Avancats
(ICREA), Barcelona, Spain\\
$^3$King Abdulaziz University, Jeddah, 22254, Saudi Arabia\\
 $^4$Skobeltsyn Institute of Nuclear Physics, Lomonosov
Moscow State University, Leninskie Gory, GSP-1, 119991, Moscow, Russia}

\begin{abstract}
The possibility to construct inflationary models for the renormalization-group improved potentials corresponding to scalar electrodynamics and to $SU(2)$ and $SU(5)$ models is investigated. In all cases, the tree-level potential, which corresponds to the cosmological constant in the Einstein frame, is seen to be non-suitable for inflation. Rather than adding the Hilbert--Einstein term to the action, quantum corrections to the potential, coming from to the RG-equation, are included. The inflationary scenario is analyzed with unstable de Sitter solutions which correspond to positive values of the coupling function, only. We show that, for the finite $SU(2)$ model and $SU(2)$ gauge model, there are no de Sitter solutions suitable for inflation,  unless exit from it occurs according to some weird, non-standard scenarios. Inflation is realized both for scalar electrodynamics and for $SU(5)$ RG-improved potentials, and the corresponding values of the coupling function are seen to be positive. It is shown that, for quite reasonable values of the parameters, the inflationary models obtained both from scalar electrodynamics and from the $SU(5)$ RG-improved potentials, are in good agreement with the most recent observational data coming from the Planck2013 and BICEP2 collaborations.
\end{abstract}


\preprint{arXiv:1408.1285}
\maketitle

\section{Introduction}

Precise astronomical data coming from recent
observational missions~\cite{WMAP,Planck2013,BICEP2} (see
also~\cite{Seljak}) support the existence of an extremely short
and intense stage of accelerated expansion in the early Universe
(inflation), as well as of a long-lasting accelerated phase at
present. These results set important restrictions on existing inflationary
models~\cite{Starobinsky:1979ty,Mukhanov:1981xt,Guth:1980zm,
Linde:1981mu,SU5inflation,Albrecht:1982wi,
inflation2,nonmin-infl,Salopek,OdintsovNOnmin,nonmin-quant,Cerioni,HiggsInflation,
DeSimone:2008ei,GB2013,KL2013}
(see also~\cite{Lindebook,19,Inflation_review} and references
therein).

Moreover, these observational data give strong support to the fact that
the post-inflationary Universe was nearly homogeneous, isotropic
and spatially flat, at very large distances or short times.
Presently, the evolution of our Universe can be well described
in terms of a spatially flat
Friedmann--Lema\^{i}tre--Robertson--Walker (FLRW) background and
cosmological perturbations and models with scalar fields are
very well suited to describe an evolution of this kind. It has also
been proven that some modified gravity models, as $f(R)$
gravity, can in a sense be considered as generic General Relativity
models with additional scalar fields. This is the reason why
scalar fields play such an essential role in modern cosmology;
in particular, in the current description of the evolution of
the Universe at a very early
epoch~\cite{Starobinsky:1979ty,Mukhanov:1981xt,Guth:1980zm,Linde:1981mu,
SU5inflation,Albrecht:1982wi,inflation2}.
Many inflationary models involve scalar
fields nonminimally coupled to the Ricci curvature scalar~\cite{nonmin-infl,OdintsovNOnmin,nonmin-quant,HiggsInflation,Cerioni,DeSimone:2008ei,
GB2013,KL2013}.
Note, however, that predictions of the simplest inflationary models with minimal couplings to scalar fields, as the $\lambda\phi^4$ model, are actually in sharp disagreement with the Planck2013
results~\cite{Planck2013}, and that some of these inflationary scenarios had to be improved by adding a tiny nonminimal coupling of the inflaton field to gravity~\cite{GB2013,KL2013}.
The conditions for a model to be consistent with the BICEP2
result have been examined in many papers already (see, e.g.,\cite{Wan:2014fra,Hossain:2014ova, Garcia-Bellido:2014eva, Barranco:2014ira, Martin:2014lra, Gao:2014pca, CH-CHZ, Hamada:2014xka, KS-KSY-HKSY,Inagaki:2014wva}). And it is in fact possible to reconstruct models with minimally coupled scalar fields which realize an inflation compatible with the Planck and the BICEP2 results, by using, e.g., the algorithm proposed in~\cite{Bamba:2014daa}.

Also a very crucial issue is the possibility to describe inflation
using particle physics models~\cite{Cervantes-Cota1995,Lyth:1998xn}, as
the Standard Model of elementary
particles~\cite{HiggsInflation,DeSimone:2008ei} or some other
Quantum Field Theory, as supersymmetric
models~\cite{SUSEinflation} or non-supersymmetric grand unified
theories (GUTs)~\cite{Albrecht:1982wi,GUT_Inflation}. This is a fundamental step towards the
longstanding and very ambitious program of the unification of physics at all scales.

As a very important step towards this goal, one should not forget to take into account
 quantum effects of quantum field theories in curved
space-time at the inflationary epoch (see~\cite{BOS} for a
general introduction). It is well understood that quantum GUTs
in curved space-time lead also to curvature induced phase transitions (for
a complete description, see~\cite{BOS,BO1985,Elizalde:1993ee}).
Note moreover that curvature induced phase transitions, as discussed in~\cite{BOS,BO1985}, may be described with better accuracy when one considers this phenomena within renormalization-group improved effective potentials (see~\cite{Elizalde:1993ee}). Indeed, in this case, the summation of all leading logs is done and the corresponding RG-improved effective potential goes far beyond the one-loop approximation. These phase transitions are very important in early-universe
cosmology. Specifically, some models of the inflationary
universe~\cite{SU5inflation,Inflation_review} are based on
first-order phase transitions, which took place during the
reheating phase of the Universe in the grand unification
epoch~\cite{Albrecht:1982wi}. Also, curved space-time effects in the
grand unification epoch
cannot be dismissed, simply considered to be negligible. Quite on the
contrary, all these theories should be treated as quantum field
theories in curved space-time, as discussed some time ago
in~\cite{Elizalde:1993ee}.
Indeed, it must be properly emphasized that the recent results by the
BICEP2 collaboration~\cite{BICEP2} point clearly towards the GUT
scale, what is a very impressive hint of a probably deep connection
of inflation with the GUT epoch and a validation of the
arguments in paper~\cite{Elizalde:1993ee}. As was
emphasized there, GUTs corresponding to the very early universe
ought to be treated as quantum field theories in curved
space-time, in a proper and rigorous way.

Anyhow, in the lack of a clear prescription for how to combine quantum
field theory at non-zero temperature and quantum field theory in
curved space-time (external temperature and external
gravitational field), it is natural to start by addressing just the
second part of this problem.
The renormalization-group improved effective potential for an
arbitrary renormalizable massless gauge theory in curved
space-time was discussed in~\cite{Elizalde:1993ee}, working in
the linear curvature approximation, because at least these
linear curvature terms ought to be taken into account in the
discussion of the effective potential corresponding to GUTs in
the early universe. Quantum corrections
with account to gravity effects are predicted to be even more
important in a chaotic inflationary model~\cite{Lindebook}. By
generalizing the Coleman--Weinberg approach corresponding to the
case of the effective potential in flat space-time, the authors found, at a first
instance, the explicit form of the
renormalization group (RG) improved effective potential in
curved space for scalar electrodynamics, the finite $SU(2)$ model,
the $SU(2)$ gauge model, and the $SU(5)$ GUT model. The possibility of
corresponding curvature-induced phase transitions was also
investigated.

By carrying out one-loop calculations in a weak gravitational field it was
shown~\cite{Freedman,BirrellDavies} that it is necessary to introduce
an induced gravity term proportional to $R\phi^2$ in order to renormalize the theory
of a scalar field in curved space-time.
Here, we consider different RG-improved effective potentials for
the tree-level potential $\lambda \phi^4- \xi\phi^2R$.
These potentials were proposed in~\cite{Elizalde:1993ee,Elizalde:1994im}.
In the Einstein frame the tree-level potential corresponds to the cosmological
constant and is not suitable for the construction of an inflationary scenario.
We will check the possibility to construct inflationary models using the RG-improved effective potentials and consider inflation based on an unstable de
Sitter solution. We will start by checking the existence of such solutions.
Then we will examine if the inflationary model with this potential is compatible with the
Planck2013 and BICEP2 data. To do that, we will use conformal transformation and the
slow-roll parameters in the Einstein frame.

The paper is organized as follows. In Sec.~II, we consider the action with
a nonminimally coupled scalar field and the corresponding equations of motion.
In Sec.~III we summarize the standard theory of Lyapunov's stability, as applied to de Sitter solutions in these models. In Sec.~IV, we discuss the general procedure for the construction of RG-improved effective potentials. The existence and stability of de Sitter solutions in scalar electrodynamics is considered in Sec.~V. Sections~VI and VII are devoted to RG-improved effective potentials for the cases of the finite  $SU(2)$  and of the $SU(2)$ models, respectively.
Unstable de Sitter solutions for the $SU(5)$ model are dealt with in Sec.~VIII.
In Sec.~IX, cosmological parameters from the inflationary models considered are extracted, and it is shown that, for some specific models, they are compatible with the Planck13 and BICEP2 results.
The last section is devoted to conclusions.

\section{Models with nonminimally coupled scalar fields}
Different models with the Ricci scalar multiplied by a function
of the scalar field are being intensively studied in
cosmology~\cite{nonmin-infl,Cerioni,HiggsInflation,Cooper:1982du,Kaiser,KKhT,Polarski,Elizalde,
Kamenshchik:2012rs,CervantesCota:2010cb,KTV2011,Sami:2012uh,ABGV}
(see also~\cite{Book-Capozziello-Faraoni,Fujii_Maeda,NO-rev} and
references therein).
Generically, these models are described by
\begin{equation}
\label{action}
S=\int d^4 x \sqrt{-g}\left[
U(\phi)R-\frac12g^{\mu\nu}\phi_{,\mu}\phi_{,\nu}-V(\phi)\right],
\end{equation}
where $U(\phi)$ and $V(\phi)$ are differentiable functions of
the scalar field $\phi$, $g$ is the determinant of the metric
tensor
$g_{\mu\nu}$, and $R$ the scalar curvature. We will use the signature
$(-,+,+,+)$ throughout.

Let us consider a spatially flat FLRW universe with metric
interval
\begin{equation*}
ds^2={}-dt^2+a^2(t)\left(dx_1^2+dx_2^2+dx_3^2\right).
\end{equation*}
The Friedmann equations, derived by variation of  action~(\ref{action}), have the following form~\cite{KTV2011}:
\begin{equation}
\label{Fr1}
6UH^2+6\dot U H=\frac{1}{2}\dot\phi^2+V,
\end{equation}
\begin{equation}
\label{Fr2}
2U\left(2\dot H+3H^2\right)+4\dot U H+2\ddot U={}-
\frac{1}{2}\dot\phi^2+V,
\end{equation}
where the Hubble parameter is the logarithmic derivative of the
scale factor:
$H=\dot a/a$ and differentiation with respect to time $t$ is denoted by a dot.
Variation of the action (\ref{action}) with respect to $\phi$
yields
\begin{equation}
\label{Fieldequ}
\ddot \phi+3H\dot\phi+V^{\prime}=6\left(\dot H
+2H^2\right)U^{\prime}\,,
\end{equation}
where the prime denotes derivation with respect to
the argument of the functions, that is, the scalar
field $\phi$.
 Combining  Eqs.~(\ref{Fr1}) and (\ref{Fr2}), we obtain
\begin{equation}
\label{Fr21}
4U\dot H-2\dot U H+2\ddot U +\dot\phi^2=0.
\end{equation}
From Eqs.~(\ref{Fr1})-(\ref{Fr21}), one can get the following
system of first order differential equations~\cite{ABGV}:
\begin{equation}
\begin{split}
  \dot\phi&=\psi,\\
\dot\psi&={}-3H\psi-\frac{\left[(6
U''+1)\psi^2-4V\right]U'+2UV'}{2\left(3 {U'}^2+ U\right)},\\
\dot
H&={}-\frac{2U''+1}{4\left(3{U'}^2+U\right)}\psi^2+\frac{2U'}{{3{U'}^2+U}}H\psi
-\frac{6{U'}^2}{3{U'}^2+U}H^2+\frac{U'V'}{2\left(3{U'}^2+U\right)}\,.
\end{split}
\label{FOSEQU}
\end{equation}
Note that Eq.~(\ref{Fr1}) is not a consequence of the system~(\ref{FOSEQU}). On the other hand,
if Eq.~(\ref{Fr1}) is satisfied for an initial time, then it
follows from  the system (\ref{FOSEQU}),
that Eq.~(\ref{Fr1}) is also satisfied for any value of time. In
other words, it turns out that the system (\ref{FOSEQU})
is equivalent to the initial system of equations,
(\ref{Fr1})-(\ref{Fieldequ}), if and only if
one chooses the initial data so that Eq.~(\ref{Fr1}) is fulfilled.

\section{Lyapunov stability of the de Sitter solutions}

We are here considering the possibility of inflationary scenarios
in models with RG-improved potentials.
Our first goal, therefore, is to find unstable de Sitter solutions.
The standard way to explore an inflationary model is to formulate it in the Einstein frame.
This is actually very convenient when $U$ is a simple function, for
instance, for induced gravity models~\cite{Kaiser}. However, in our case the Jordan frame is more suitable to perform an analysis of the stability of the de Sitter solutions, because the potential can be expressed in terms of elementary functions in this frame only.  We will consider
the de Sitter solutions which correspond to a constant
$\phi$, only. In other words, we consider a fixed point of Eqs.~(\ref{FOSEQU}), with the additional condition~(\ref{Fr1}).

Substituting constant values for $H=H_f$ and $\phi=\phi_f$ into
Eqs.~(\ref{Fr1}) and (\ref{Fieldequ}), we get
\begin{equation}
\label{Hf}
H_f^2=\frac{V(\phi_f)}{6U(\phi_f)},
\end{equation}
\begin{equation}
V^{\prime}(\phi_f)=12H_f^2U^{\prime}(\phi_f).\label{2}
\end{equation}
Therefore, we come up with the following simple condition
\begin{equation}
\label{equphif}
2\frac{U^\prime(\phi_f)}{U(\phi_f)}=\frac{V^\prime(\phi_f)}{V(\phi_f)}.
\end{equation}

We consider the stability with respect to homogeneous isotropic
perturbations. In other words, we use (\ref{FOSEQU}) and
analyze the Lyapunov stability of the de Sitter solutions
derived from it. For this we apply Lyapunov's theorem~\cite{Lyapunov,Pontryagin} and study the
corresponding linearized system. We expand around the fixed point, in the way
\begin{equation}
\label{expanding}
\phi(t)=\phi_f+\varepsilon\phi_1(t),\qquad
\psi(t)=\varepsilon\psi_1(t),\qquad H(t)=H_f+\varepsilon H_1(t),
\end{equation}
where $\varepsilon$ is a small parameter.
Substituting (\ref{expanding}) into~\eqref{FOSEQU}, to first order in $\varepsilon$ we obtain
the following linear system\footnote{In the case of induced gravity ($U=\xi\phi^2$) a
similar stability analysis of de Sitter solutions has been
carried out in~\cite{PV2014}.}:
\begin{equation}
\label{linsystem}
\begin{split}
\dot{\phi}_1&=\psi_1,\\
\dot{\psi}_1&=\frac{V_f^\prime
U_f^\prime+2V_fU_f^{\prime\prime}-U_fV_f^{\prime\prime}}{3(U_f^\prime)^2+U_f}\phi_1-3H_f\psi_1,\\
\dot{H}_1&=\frac{(U_f^\prime V_f^{\prime\prime}-V_f^\prime
U_f^{\prime\prime})}{2(3(U_f^\prime)^2+U_f)}\phi_1+
\frac{2H_fU_f^\prime}{3(U_f^\prime)^2+U_f}\psi_1-\frac{12H_f(U_f^\prime)^2}{3(U_f^\prime)^2+U_f}H_1.
\end{split}
\end{equation}

The following matrix, $A$,  corresponds to (\ref{linsystem}):
\begin{equation}
A=    \begin{array}{||ccc||}
      0 & 1 & 0 \\
\frac{V_f^\prime
U_f^\prime+2V_fU_f^{\prime\prime}-U_fV_f^{\prime\prime}}{3(U_f^\prime)^2+U_f}
& -3H_f & 0 \\
\frac{U_f^\prime V_f^{\prime\prime}-V_f^\prime
U_f^{\prime\prime}}{2(3(U_f^\prime)^2+U_f)} &
\frac{2H_fU_f^\prime}{3(U_f^\prime)^2+U_f} &
-\frac{12H_f(U_f^\prime)^2}{3(U_f^\prime)^2+U_f} \\
    \end{array}\, .
\end{equation}
Its associated characteristic equation,
\begin{equation}
\label{lambdaequ}
\det(A-\tilde{\lambda}
I)=\left(\frac{12H_fU_f^\prime}{3(U_f^\prime)^2+U_f}+\tilde{\lambda}\right)
\left(\tilde{\lambda}(3H_f+\tilde{\lambda})
-\frac{V_f^\prime U_f^\prime+2V_f
U_f^{\prime\prime}-U_fV_f^{\prime\prime}}{3(U_f^\prime)^2+U_f}\right)=0,
\end{equation}
has the following roots:
\begin{equation}
\label{lambda123}
\tilde{\lambda}_{\pm}={}-\frac{3H_f}{2}\pm\sqrt{\frac{9H_f^2}{4}+\frac{V_f^\prime
U_f^\prime+2V_fU_f^{\prime\prime}-U_fV_f^{\prime\prime}}{3(U_f^\prime)^2+U_f}},
\qquad
\tilde{\lambda}_3={}-\frac{12H_fU_f^\prime}{3(U_f^\prime)^2+U_f}.
\end{equation}

Lyapunov's theorem~\cite{Lyapunov,Pontryagin} states that in
order to prove the stability of a fixed point of a nonlinear system it
is sufficient to prove the stability of this fixed point for the
corresponding linearized system. Stability of the linear system
relies, on its turn, on the real parts of the roots $\tilde{\lambda}_k$
of the characteristic equation (\ref{lambdaequ}), which must all
be negative. If at least one of them is positive, then the fixed
point is unstable.

To describe inflation we are interested in finding unstable de
Sitter solutions with $H_f>0$. Note that the perturbation $H_1(t)$
is not independent, because it is connected with
$\phi_1$ and $\psi_1$ due to Eq.~(\ref{Fr1}). So, the de Sitter solution is stable if
the real parts of $\tilde{\lambda}_{\pm}<0$. The real part of $\tilde{\lambda}_-$ is
always negative, hence, just $\tilde{\lambda}_+$ defines the stability.

Introducing
\begin{equation}
\label{K}
K_f\equiv\frac{V_f^\prime
U_f^\prime+2V_fU_f^{\prime\prime}-U_fV_f^{\prime\prime}}{3(U_f^\prime)^2+U_f}
=\frac{2\left(\frac{U_f^\prime}{U_f}\right)^\prime
-\left(\frac{V_f^\prime}{V_f}\right)^\prime}{\frac34\left(\frac{V_f^\prime}{V_f}\right)^2
\frac{U_f}{V_f}+\frac{1}{V_f}},
\end{equation}
we can then formulate a sufficient stability condition as follows:
the de Sitter solution ($H_f>0$) is stable at $K_f<0$ and unstable at $K_f>0$.

\section{Renormalization-group improved effective potential}

The renormalization-group improved effective potential for an
arbitrary renormalizable massless gauge theory in curved
space-time was discussed in detail in~\cite{Elizalde:1993ee}. In
this section we will just remind the reader of the basic steps for the
construction of the renormalization-group improved effective potential.

The tree-level potential reads as follows~\cite{Elizalde:1993ee}
\begin{equation}
\label{W0}
W^{(0)}(\phi)= a\lambda \phi^4-b\xi\phi^2 R =V_0-U_0R,
\end{equation}
where $a$ and $b$ are positive constants and $\xi$ is the conformal coupling.
The potential $W^{(0)}$ includes both the potential
$V_0$ and the function $U_0$ multiplied by the scalar curvature.

As is known, see~\cite{Elizalde:1993ee,BOS}, the
renormalization-group equation for the effective potential in
curved space-time has the form
\begin{equation}
\left( \mu \frac{\partial}{\partial \mu} +\beta_{\wt{g}}
\frac{\partial}{\partial \wt{g}} +\delta
\frac{\partial}{\partial
\alpha} + \beta_\xi \frac{\partial}{\partial \xi}
-\gamma \phi \frac{\partial}{\partial \phi}  \right)
W=0,
\label{RG_equ}
\end{equation}
where $\alpha$ is the gauge parameter and $\tilde g$  is the set of all coupling constants of the theory (Higgs, gauge and Yukawa ones). The standard flat-space renormalization-group
 equation~\cite{Coleman,Sher} is modified in curved space-time, for instance, it has
 an additional term related with the contribution from the nonminimal
 coupling constant $\xi$ and the corresponding $\beta_\xi$ function~\footnote{Note that, in the case of the derivation of the renormalization-group improved effective action, the corresponding RG equation~(\ref{RG_equ}) must be generalized with account also to the relevant couplings in the vacuum sector (the higher derivative gravitational terms), see e.g. the second paper of Ref.~\cite{Elizalde:1993ee}.}.

It is natural to split $W$ into two parts, namely
\begin{equation}
W\equiv V - UR \equiv af_1(p,\phi,\mu)
\phi^4 -b f_2(p,\phi,\mu)\phi^2R,
\end{equation}
where $f_1$ and $f_2$ are some unknown functions, and
$p=\{\wt{g}, \alpha, \xi \}$.
Actually, in~\cite{Elizalde:1993ee} the authors imposed the
additional restriction that, not only the function $W$ satisfies
(\ref{RG_equ}), but also that the functions $V$ and
$U$ satisfy it, separately.

It is easy to see~\cite{KTV2011} that, for
\begin{equation}
V_0(\phi) = C U_0^2(\phi),
\label{constant1}
\end{equation}
where $C$ is a constant and the model considered has a de Sitter
solution, with an arbitrary constant $\phi=\phi_f$.
Therefore, for the tree-level potential $W^{(0)}$, de Sitter
solutions do exist, for any value of $\phi_f$, and
the corresponding Hubble parameter reads
\begin{equation*}
H_{0f}=\pm \sqrt{{}-\frac{a\lambda}{6b\xi}}\phi_f.
\end{equation*}
One aim of this paper will be to consider de Sitter solutions in
cosmological models with different RG-improved $W$ potentials
and the possibility of inflationary scenarios in such models, too.

\section{Effective potentials for scalar electrodynamics}

Let us now consider the de Sitter solution for the case of the following
effective potentials for scalar electrodynamics
\begin{equation}
\label{VUSCED}
V=\frac{\lambda\phi^4}{4!}+\frac{3e^4\phi^4}{(8\pi)^2}\ln\frac{\phi^2}{\mu^2},
\qquad
U=\frac{\xi\phi^2}{2}+\frac{e^2\phi^2}{(8\pi)^2}\ln\frac{\phi^2}{\mu^2},
\end{equation}
where $e$ is a constant. Using
\begin{equation*}
\frac{U^\prime}{U}=\frac{2}{\phi}\left(1+\frac{e^2}{32\xi+e^2\displaystyle
\ln\frac{\phi^2}{\mu^2}}\right),
\qquad
\frac{V^\prime}{V}=\frac{2}{\phi}\left(2+\frac{9 e^4}{8\lambda\pi^2+9e^4\displaystyle\ln\frac{\phi^2}{\mu^2}}\right),
\end{equation*}
and the condition (\ref{equphif}), we get
\begin{equation}
\phi_f=\pm\mu\exp\left[\frac{8\pi^2}{9
e^4}\left(18e^2\xi-\lambda\right)\right].
\end{equation}
From (\ref{Hf}), we obtain
\begin{equation}
H_f^2= \frac{\mu^2e^2}{4}\exp\left[\frac{16\pi^2}{9e^4}(18\xi
e^2-
\lambda)\right]=\frac{e^2\phi_f^2}{4}.
\end{equation}
For the de Sitter solutions obtained, we then get
\begin{equation}
V_f = \frac{36e^2\xi-\lambda}{24}\phi_f^4,\qquad U_f = \frac{36e^2\xi-\lambda}{36e^2}\phi_f^2.
\end{equation}

Let us now consider the stability of the above solutions:
\begin{equation*}
2\left(\frac{U^\prime}{U}\right)^\prime|_{\phi=\phi_f}
-\left(\frac{V^\prime}{V}\right)^\prime|_{\phi=\phi_f}=
\frac{81e^8}{32\pi^4(36e^2\xi-\lambda)^2\phi_f^2}>0.
\end{equation*}
Using (\ref{K}), we get that $K_f>0$ at $U_f>0$. This means that $\tilde{\lambda}_+>0$ and that the corresponding de Sitter solution is unstable. Thus, we get in the end an unstable de Sitter solution with $H_f>0$ and $U_f>0$. Note that the variation of the Hubble parameter is considered as a function of the variations of both the scalar field and its first derivative.

\section{Finite $SU(2)$ models}

A number of grand unified theories (GUTs) turn out to yield finite
models. Some of them, as for instance the finite
supersymmetric $SU(5)$ GUT~\cite{Odintsov7}, may lead to reasonable
phenomenological consequences and deserve attention as
realistic models of grand unification. Asymptotically finite
GUTs, which are generalizations of the concept of a finite theory, have
been proposed in~\cite{Ermushev8}. In these theories, the zero
charge problem is absent, both in the UV and in the IR limits, since in these
limits the effective coupling constants tend to some constant values
(corresponding to finite phases).

When we consider flat space-time there is not much sense in
discussing quantum corrections to the classical potential, in a massless
finite or massless asymptotically finite GUT, since  they either
are simply absent or highly suppressed asymptotically.
However, when we study finite theories in curved space-time~\cite{9}
(for a general review, see~\cite{BOS}) the situation changes
drastically~\cite{Elizalde:1994im}.

In the following, we will study de Sitter solutions in cosmological models with
renormalization-group improved effective $SU(2)$ potentials for the two finite
theories in curved space-time constructed in~\cite{Elizalde:1994im}.
In those models the coupling parameter corresponding to the nonminimal scalar-gravitational
interaction $\xi$ depends on $\vartheta$, where $\vartheta=\frac{1}{2} \ln (\phi^2/\mu^2)$.

The general structure of the one-loop effective coupling
constant $\xi (\vartheta)$ for ``finite'' theories in curved space-time
has been obtained in~\cite{9}:
\begin{equation}
\label{xi}
\xi (\vartheta) = \frac{1}{6} + \left( \xi_0 - \frac{1}{6} \right)
\exp
(Cg^2\vartheta),
\end{equation}
with constant $\xi_0$ and  $C\neq 0$.

In particular, for the $SU(2)$ finite gauge model~\cite{4},
it was obtained that $C=6$ or $C\simeq 28$~\cite{9}. Hence, in such theories
we have $|\xi (\vartheta)| \rightarrow
\infty$ (non-asymptotical conformal invariance)
in the UV limit ($t\rightarrow \infty$). In the models
which have $C<0$ one gets $\xi (\vartheta) \rightarrow
1/6$ (asymptotical conformal invariance).

The tree-level potential is taken to be of the form \eqref{W0}
and the RG-improved potential reads as follows (see~\cite{Elizalde:1994im} for
details)
\begin{equation}
W = a\lambda (\vartheta) f^4(\vartheta) \phi^4 -b\xi
(\vartheta) f^2(\vartheta) \phi^2 R\,.
\label{VSU2improve}
\end{equation}
 Notice that this potential is actually obtained in the linear
curvature approximation, what is good enough for GUTs
corresponding to the curved space-time corresponding to the early universe
\cite{Elizalde:1994im}.

The function $f(\vartheta)$ is defined as~\cite{Elizalde:1994im}
\begin{equation}
f(\vartheta) = \exp \left[ - \int\limits_0^\vartheta d\vartheta' \,
\bar{\gamma} \left( \wt{g}
(\vartheta'), \alpha (\vartheta') \right) \right].
\label{6}
\end{equation}
Therefore, the form of the RG-improved effective potential is
determined by
the $\bar{\gamma}$-function of the scalar field in
(\ref{6}). At the one-loop level, $\bar{\gamma}(\vartheta) \sim a_1
g^2(\vartheta) +a_2 h^2(\vartheta)$,
where $a_1$ and $a_2$ are constants, with values which depend on
the choice of gauge and on other features of the theory. As $h^2=
\kappa_1 g^2$, it turns out that $\bar{\gamma}(\vartheta) \sim (a_1
+\kappa_1 a_2)g^2(\vartheta)$. Through the choice of the gauge parameter,
one can obtain different values for $\bar{\gamma}$. To reach as much
`finiteness' in our theory as possible, we can choose a gauge such
that the one-loop $\bar{\gamma}$-function be equal to zero.
This choice is always possible; moreover, in supersymmetric
finite theories it does appear in a very natural way (specially if the
superfield technique is used).

After having done all this, it turns out that the RG-improved effective
potential (in the linear-curvature and leading-log approximation)
for a `finite' theory in curved space-time is given by
\begin{equation}
W= a\kappa_1 g^2 \phi^4 - b \xi (\vartheta)  \phi^2R\,.
\label{9}
\end{equation}
A straightforward calculations show that a de Sitter solution exist at $C=0$, only.
It corresponds to $\xi(\vartheta)=\xi_0$ and it is not interesting, because it leads
to $W^{(0)}(\phi)$, given by \eqref{W0}.

Another possibility is to keep the gauge arbitrary; then we
cannot demand that $\bar{\gamma}$ vanishes. We get in this case
\begin{equation}
W= a\kappa_1 g^2 f^4(\vartheta) \phi^4 - b \xi
(\vartheta)f^2(\vartheta)  \phi^2 R
\label{10}
\end{equation}
and $\bar{\gamma} = C_1 g^2$, $C_1$ being some constant which depends
on the gauge parameter and on the features of the theory,
$f(\vartheta)=
\exp (-C_1 g^2 \vartheta)$, and $\xi (\vartheta)$ is given by
(\ref{xi}).

From
\begin{equation}
\label{VUmodi}
V= a\kappa_1 g^2 f^4(\vartheta) \phi^4,
\qquad
U= b \xi (\vartheta)f^2(\vartheta)\phi^2,
\end{equation}
where
\begin{equation*}
f(\vartheta)=\left(\frac{\phi}{\mu}\right)^{-C_1g^2},\qquad
\xi(\vartheta)=\frac16+\left(\xi_0-\frac16\right)\left(\frac{\phi}{\mu}\right)^{Cg^2},
\end{equation*}
we get
\begin{equation}
V=ak_1g^2\left(\frac{\phi}{\mu}\right)^{-4C_1g^2}\phi^4,\qquad
U=b\left(\frac16+\left(\xi_0-\frac16\right)\left(\frac{\phi}{\mu}\right)^{Cg^2}\right)
\left(\frac{\phi}{\mu}\right)^{-2C_1g^2}\phi^2,
\end{equation}
and
 \begin{equation}
\frac{U^\prime}{U}=\frac{2(1-C_1g^2)}{\phi}+\frac{(6\xi_0-1)Cg^2\left(\frac{\phi^2}
{\mu^2}\right)^{Cg^2/2}}
{\left[1+(6\xi_0-1)\left(\frac{\phi^2}{\mu^2}\right)^{Cg^2/2}\right]\phi},
\qquad
\frac{V^\prime}{V}=\frac{4(1-C_1g^2)}{\phi}\,.
\end{equation}
Now, we use the de Sitter condition (\ref{equphif})
and conclude that this equation has as only solution $C=0$.

Summing up, in both cases there is no de Sitter solution for a nonconstant $\xi$.
In the case of a constant $\xi$, we get a model with a power-law potential $V$ and power-law coupling function $U$ which satisfies the condition (\ref{constant1}). Thus, if the signs of the constants are such that
\begin{equation}
H^2_f = \frac{k_1g^2}{6b\xi_0}\phi_f^2\left(\frac{\phi_f^2}{\mu^2}\right)^{-C_1g^2}>0,
\end{equation}
and then de Sitter solutions do exist for any constant value of $\phi_f$. Note that, after conformal transformation to the Einstein frame, one gets a model with a minimally coupled scalar field, whose potential is a constant. Namely, in the case $C=0$ we get $\xi=\xi_0$ and, using (\ref{VUmodi}),
\begin{equation*}
V_\mathrm{E} = \frac{a\kappa_1g^2}{4\xi_0^2\kappa^4b^2}.
\end{equation*}
See formula (\ref{poten}) in Section IX. Note that above property holds in the presence of quantum corrections as we take them into account via the effective potential.

\section{The $SU(2)$ gauge model}

Let us consider the cosmological model with
\begin{equation}
V=\frac{k_1\phi^4f^4(\vartheta)g^2(\vartheta)}{24} ,
\qquad
 U=\frac{\phi^2f^2(\vartheta)}{12}
 \left[1+\left(6\xi-1\right)\tilde{\Theta}^{{}-k_3}\right],
\end{equation}
where we use $k_3=(12-5k_1/3-8k_2)/\tilde{a}^2$ to
simplify notations. $k_i$ and $\tilde{a}$ are constant. Also,
\begin{equation*}
\tilde{\Theta}=1+\frac{\tilde{a}^2g^2\vartheta}{(4\pi)^2},\qquad
f(\vartheta)=\tilde{\Theta}^{(6-4k_2)/\tilde{a}^2},\qquad
g^2(\vartheta)=\frac{g^2}{\tilde{\Theta}}.
\end{equation*}
By straightforward calculation, we obtain
\begin{equation*}
\frac{V^\prime}{V}=\frac{4}{\phi}+\left\{\frac{4(6-4k_2)}{\tilde{a}^2}-1\right\}
\frac{1}{\tilde{\Theta}}\frac{d\tilde{\Theta}}{d\phi}\,,
\qquad
\frac{U^\prime}{U}=\frac{2}{\phi}+\left\{\frac{2(6-4k_2)}{\tilde{a}^2}
-\frac{\left(6\xi-1\right)
k_3}{{\tilde{\Theta}}^{k_3}+6\xi-1}\right\}\frac{1}{\tilde{\Theta}}\frac{d\tilde{\Theta}}{d\phi}\,.
\end{equation*}
Then, using (\ref{equphif}) and the conditions
  $\tilde{\Theta}_f^{k_3}\neq-6\xi+1\,$, $\frac{d\tilde{\Theta}}{d\phi}\neq 0$,
 we get
\begin{equation}
\tilde{\Theta}_f^{k_3} =\left(2k_3-1\right)\left(6\xi-1\right).
\end{equation}
Therefore, there is no de Sitter solution for $\xi=1/6$. For other values of $\xi$, we get
\begin{equation}
\label{Theta_f}
\tilde{\Theta}_f=\left[\left(6\xi-1\right)\left(2k_3-1\right)
\right]^{1/k_3}
\quad\mbox{and}\quad
\phi_f=\pm\mu\exp\left(\frac{(4\pi)^2}{\tilde{a}^2g^2}
\left[\left[\left(6\xi-1\right)\left(2k_3-1\right) \right]
^{1/k_3}-1\right]\right).
\end{equation}

Note that $\tilde{\Theta}_f\neq 0$, hence $k_3\neq 1/2$.
Using (\ref{Hf}), we calculate the Hubble parameter $H$ for the de Sitter
solution
\begin{equation}
H^2_f=\frac{g^2k_1(2k_3-1)}
{24k_3}\phi_f^2\tilde{\Theta}_f^{\frac{2(6-4k_2)}{\tilde{a}^2}-1}\,.
\end{equation}

In~\cite{SU2} the authors show that asymptotically free models
exist only when $f^2(\vartheta)\sim g^4(\vartheta)$, what
corresponds to $\tilde{a}^2=4k_2-6$. For this choice of the parameter $\tilde{a}^2$, we obtain
\begin{equation*}
V_f=\frac{g^2k_1\phi_f^4}{24}\tilde{\Theta}_f^{-5},\qquad
U_f=\frac{\phi_f^2}{12}\tilde{\Theta}_f^{-2} \left[1+\left(6\xi-1\right)
\tilde{\Theta}_f^{-k_3}\right]=\frac{\phi_f^2k_3}{6(2k_3-1)\tilde{\Theta}_f^2}\,.
\end{equation*}
Note that $k_3=-2-5k_1/(3(4k_2-6))=-2-5k_1/(3\tilde{a}^2)$. Therefore,
\begin{equation}
H^2_f={}-\frac{g^2k_1(1-2k_3)}{24k_3}\phi_f^2\tilde{\Theta}_f^{-3}
=\frac{5k_1(2k_1+3\tilde{a}^2)}{24(5k_1+6\tilde{a}^2)}g^2\phi_f^2
\left[\left(2k_3-1\right)\left(6\xi-1\right)
\right]^{-3/k_3}.
\end{equation}
From here, we get that $H^2_f>0$ provided either $-3\tilde{a}^2/2<k_1<-6\tilde{a}^2/5$ that is equivalent to $0<k_3<1/2$, or $0<k_1<\infty$ that correspond to $-\infty<k_3<0$.

Let us consider the stability of the de Sitter solutions in the case $V_f>0$ and $U_f>0$.
In this case, the sign of $K_f$ coincides with the sign of
\begin{equation*}
2\left(\frac{U^\prime}{U}\right)^\prime|_{\phi=\phi_f}
-\left(\frac{V^\prime}{V}\right)^\prime|_{\phi=\phi_f}
=\frac{\tilde{a}^4g^4(2k_3-1)}{512\pi^4\phi_f^2\tilde{\Theta}_f^2}\,.
\end{equation*}
De Sitter solutions exist only for $k_3<1/2$. The additional condition  $U_f>0$ gives  $k_3<0$.
We can see that $K_f<0$ at any $k_3<0$. Thus, for this model with $U_f>0$ we have  stable de Sitter solutions~only.

\section{The $SU(5)$ RG-improved potential}

Now we study the RG-improved potential for the $SU(5)$ GUT \cite{8}.
In flat space this theory has been used for the discussion of
inflationary cosmology~\cite{Lindebook,SU5inflation}.
We assume that the breaking $SU(5)$ $\rightarrow$ $SU(3) \times
SU(2) \times U(1)$ has taken place.

The RG-improved effective potential has the following form:
\begin{equation}
\label{SU5pot}
V = \frac{3375}{512} \left(g^2 - \frac{g^2}{f_5^{16/9}} \right)
\phi^4 f_5^4,\qquad
U= \frac{15}{4} \left[ \frac{1}{6} + \left( \xi - \frac{1}{6}\right) \breve{\Theta}^{-9/8}
\right]
\phi^2 f_5^2,
\end{equation}
where $\breve{\Theta}=1+ \frac{5g^2\vartheta}{3\pi^2}$,  $f_5= \breve{\Theta}^{9/16}$, $g$ is a nonzero constant.

To get a de Sitter solution, we use Eq.~(\ref{equphif}), which for
the model at hand reads
\begin{equation}
\label{Equtheta}
\breve{\Theta}'\frac{(9\breve{\Theta}-5)(6\xi-1)+4\breve{\Theta}^{9/8}}{\breve{\Theta}(\breve{\Theta}-1)(\breve{\Theta}^{9/8}+6\xi-1)}=0.
\end{equation}
Thus, $\breve{\Theta}\neq 0$ and $\breve{\Theta}\neq 1$.
Note that
\begin{equation}
\label{DTheta}
\breve{\Theta}'=\frac{5g^2}{3\pi\phi}\neq 0.
\end{equation}

It is easy to see that, for $\xi=1/6$, there is no de Sitter solution.
For other values of $\xi$, the de Sitter solutions are defined by\begin{equation}
H^2_{ds}=\frac{225\phi_f^2\breve{\Theta}_f^{5/4}
g^2(\breve{\Theta}_f-1)}{128(\breve{\Theta}_f^{9/8}+6\xi-1)}.
\end{equation}
The number $\breve{\Theta}_f$ is a root of Eq. (\ref{Equtheta}), which can be rewritten as follows:
\begin{equation}
\label{xicond}
\xi=\frac{1}{6}-\frac{2\breve{\Theta}_f^{9/8}}{3(9\breve{\Theta}_f-5)}.
\end{equation}
We can eliminate $\xi$ and express $H_{dS}^2$ as
\begin{equation}
\label{HdS_SU5}
H_{dS}^2=\frac {225 g^2\phi^2\, \left( \breve{\Theta}_f-1
\right) {\breve{\Theta}_f}^{5/4}}{128 \left( {
\breve{\Theta}_f}^{9/8}- \frac{\left(
-9\breve{\Theta}_f+5+4{\breve{\Theta}_f}^{9/8} \right)}{\left( 9\,\breve{\Theta}_f-5 \right)}-1
\right)}=\frac{25}{128}\breve{\Theta}_f^{1/8}(9\breve{\Theta}_f-5)\phi^2_fg^2.
\end{equation}

Therefore, the Hubble parameter $H$ is real if and only if
$\breve{\Theta}_f\geqslant 5/9$,
it is possible for $\xi<1/6$ only.
Using (\ref{xicond}),   we get:
\begin{equation}
V_f=\frac{3375}{512}g^2\left(1-\breve{\Theta}^{-1}_f\right)\phi^4_f\breve{\Theta}_f^{9/4}\quad\mbox{and}\quad
U_f=\frac{15}{4}\left(\frac16-\frac{2}{3(9\breve{\Theta}_f-5)}\right)\phi_f^2\breve{\Theta}_f^{9/8}\,.
\end{equation}
We see that $U_f<0$ at $5/9<\breve{\Theta}_f<1$ and $U_f>0$ for $1<\breve{\Theta}_f$.
Let us consider the stability of the de Sitter solutions here obtained.
Note that we used the conditions  $\breve{\Theta}_f\neq 1$:
\begin{equation}
\label{Dequ}
2\left(\frac{U^\prime}{U}\right)^\prime|_{\phi=\phi_f}
-\left(\frac{V^\prime}{V}\right)^\prime|_{\phi=\phi_f}
=\frac{(5-\breve{\Theta}_f)({\breve{\Theta}'_f})^2}{8(\breve{\Theta}_f-1)^2\breve{\Theta}_f^2}\,.
\end{equation}
For $1<\breve{\Theta}_f$,  $U_f>0$ and $V_f>0$, so the denominator of $K_f$ calculated by (\ref{K}) is positive, and thus the sign of $K_f$ can be determined by the numerator that was calculated in (\ref{Dequ}). We come to the conclusion that  $K_f>0$, for $1<\breve{\Theta}_f<5$, and  $K_f<0$, for $5<\breve{\Theta}_f$.

Using (\ref{DTheta}), we get
\begin{equation}
\label{expession1}
K_f=\frac{9375\,{g}^{6} \left(5-\breve{\Theta}_f\right){\breve{\Theta}_f }^{\frac18}\phi_f^{2} \left(9\breve{\Theta}_f-5\right)}{32\left[128{\breve{\Theta}_f}^{{\frac {7}{8}}}{\pi }^{4} \left(
\breve{\Theta}_f-1 \right)\left(9\breve{\Theta}_f-5\right) +30375\left( {\frac {16}{15}}\breve{\Theta}_f{\pi }^{2} \left( \breve{\Theta}_f-1 \right) +\frac{{g}^{2} }{9}\left( 9\,\breve{\Theta}_f-5\right)\right)^{2}\right]}.
\end{equation}

Let us now return to the interval $5/9<\breve{\Theta}_f<1$. In this interval the numerator of $K_f$ is positive. The  sign of $K_f$ can be determined by the denominator
\begin{equation}
S=128{\breve{\Theta}_f}^{{\frac {7}{8}}}{\pi }^{4} \left(
\breve{\Theta}_f-1 \right)\left(9\breve{\Theta}_f-5\right) +30375\left( {\frac {16}{15}}\breve{\Theta}_f{\pi }^{2} \left( \breve{\Theta}_f-1 \right) +\frac{{g}^{2} }{9}\left( 9\,\breve{\Theta}_f-5\right)\right)^{2}.
\end{equation}
\begin{figure}[!h]
\centering
\includegraphics[width=10cm]{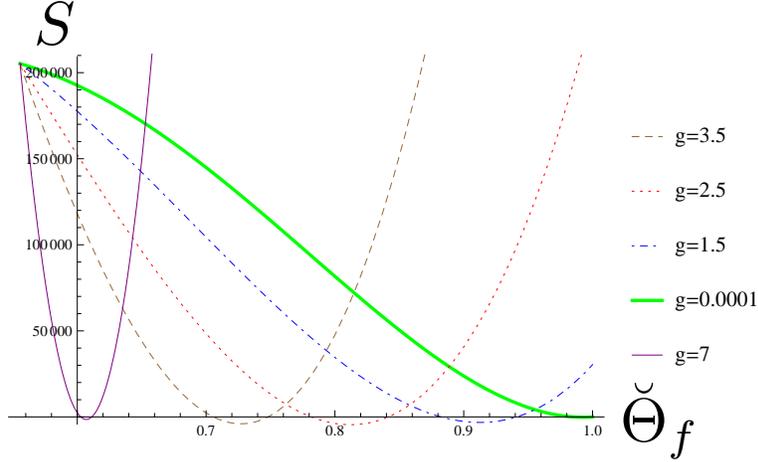}\  \  \  \
\caption{The function $S(\breve{\Theta}_f)$ for different values of $g$.}
\label{ST}
\end{figure}
The denominator of $K_f$ (the function $S$) is plotted in Fig.~\ref{ST}
for different values of $g$. One can see that the sign of this denominator depends on the value of $\breve{\Theta}_f$.
So, for  $U_f<0$ unstable de Sitter solutions can exist.

\section{Inflationary model consistent with observational results}

\subsection{Parameters of an inflationary model}
Our goal is to construct an inflationary model using the RG-improved potentials and
to examine if the inflationary model with this potential is compatible with the
Planck13 and BICEP2 data.

Much of the formalism developed for calculating the parameters of inflation, for example, the primordial spectral index $n_s$, assume General Relativity models with minimally coupled scalar fields.
The standard way to use this formalism is to perform a conformal transformation and to
consider the model in the Einstein frame (see, for example~\cite{DeSimone:2008ei}). It has been shown~\cite{Kaiser}, that in the case of quasi de Sitter expansion there is no difference between
spectral indexes calculated either in the Jordan frame directly, or in the Einstein frame after conformal transformation.

Let us make the conformal transformation of the metric
\begin{equation*}
\tilde{g}_{\mu\nu} = 2 \kappa^2 U(\phi) g_{\mu\nu},
\end{equation*}
where quantities in the new frame are marked with a tilde, and the quantity
$\kappa^2 = 8\pi M_{pl}^{-2}$, where $M_{pl}$ is the Planck mass.
We also introduce a new scalar field $\varphi$, such that
\begin{equation}
\frac{d\varphi}{d\phi} = \frac{\sqrt{U+3U'^2}}{{\sqrt{2}\kappa}U}
\quad\Rightarrow\quad
\varphi = \frac{1}{\sqrt{2}\kappa}\int \frac{\sqrt{U+3U'^2}}{U} d\phi.
\label{scal1}
\end{equation}
We thus get a model with for a minimally coupled scalar field,
described by the following action:
\begin{equation}
S =\int d^4x\sqrt{-\tilde{g}}\left[\frac{1}{2\kappa^2}R(\tilde{g}) -
\frac12\tilde{g}^{\mu\nu}\varphi_{,\mu}\varphi_{,\nu}+V_\mathrm{E}(\varphi)\right],
\label{action1}
\end{equation}
where
\begin{equation}
V_\mathrm{E}(\varphi) = \frac{
V(\phi(\varphi))}{4\kappa^4U^2(\phi(\varphi))}.
\label{poten}
\end{equation}
Inflationary universe models are based upon the possibility of a slow evolution of some scalar
field $\varphi$ in the potential $V(\varphi)$.  The slow-roll approximation, which neglects the most
slowly changing terms in the equations of motion, is used.
To calculate parameters of inflation that can be tested via observations, we use the slow-roll approximation parameters of the potential.

As known~\cite{Salopek,Liddle:1994dx} (see also~\cite{Kaiser,Bamba:2014daa}), the slow-roll parameters $\epsilon$, $\eta$ and $\zeta$
are connected with the potential in the Einstein frame as follows\footnote{We use $\zeta$ to denote the
third slow-roll parameter instead of $\xi$, because $\xi$ denotes the coupling strength.}:
\begin{equation}\label{SLP}
\epsilon \equiv \frac{1}{2\kappa^2} \left(
\frac{V_{\mathrm{E}_,\varphi}'(\varphi)}{V_\mathrm{E}(\varphi)} \right)^2\, ,\qquad
\eta \equiv \frac{1}{\kappa^2}
\frac{V_{\mathrm{E}_, \varphi}''(\varphi)}{V_\mathrm{E}(\varphi)}\, , \qquad
\zeta^2 \equiv \frac{1}{\kappa^4} \frac{V_\mathrm{E}'(\varphi)
V_{\mathrm{E}_,\varphi}'''(\varphi)}{V_\mathrm{E}(\varphi)^2}\, .
\end{equation}
Note that the prime denotes derivative with respect to
the argument of the functions, that is $\varphi$, so $V_{\mathrm{E}_, \varphi}' (\varphi) \equiv
\frac{dV_\mathrm{E}(\varphi)}{d\varphi}$. We add the additional subscript $_{,\varphi}$ to denote derivatives with respect to~$\varphi$.
During inflation, each of these  parameters should remain to be less than one.

It is suitable to calculate the slow-roll parameters as functions of the initial scalar field $\phi$. It is easy to see~\cite{DeSimone:2008ei}, that
\begin{equation}
\epsilon(\phi)=\frac{1}{2\kappa^2}\left(\frac{V_\mathrm{E}'}{V_\mathrm{E}}\right)^2
\left(\frac{d\varphi}{d\phi}\right)^{-2},\quad \eta(\phi)=\frac{1}{\kappa^2}\left[
\frac{V_\mathrm{E}''}{V_\mathrm{E}}\left(\frac{d\varphi}{d\phi}\right)^{-2}
\!-\frac{V_\mathrm{E}'}{V_\mathrm{E}}\left(\frac{d\varphi}{d\phi}\right)^{-3}
\frac{d^2\varphi}{d\phi^2}
\right]\,,
\end{equation}
where the prime denotes now derivative with respect to $\phi$.
We get
\begin{equation}\label{SLP_phi}
\epsilon(\phi) = \frac{1}{2\kappa^2}\frac{{(V_\mathrm{E}')}^2}{V_\mathrm{E}^2Q}\, ,\qquad
\eta(\phi) =\frac{1}{\kappa^2V_\mathrm{E}Q}\left[V_\mathrm{E}''-\frac{V_\mathrm{E}'Q'}{2Q}\right] \, , \quad
\quad\mbox{where}\quad
Q=\frac{U+3U'^2}{2\kappa^2U^2}.
\end{equation}
Similar calculations yield
\begin{equation}
\label{zeta2}
   \zeta^2 =\frac{V'_E}{\kappa^4V_\mathrm{E}^2Q^2}\left[V_\mathrm{E}'''-\frac{3V_\mathrm{E}''Q'}{2Q}-\frac{V_\mathrm{E}'Q''}{2Q}
+\frac{V_\mathrm{E}'(Q')^2}{Q^2}\right].
\end{equation}

The number of e-foldings of a slow-roll inflation is given by the following integral~\cite{DeSimone:2008ei}:
\begin{equation}
N_e(\phi)={\kappa^2}\int\limits_{\varphi_{\mathrm{end}}}^{\varphi}
\left|\frac{V_{\mathrm{E}}(\tilde{\varphi})}{V_{\mathrm{E},\varphi}'(\tilde{\varphi})}\right|
\,d\tilde{\varphi}=\kappa^2\int\limits_{\phi_{\mathrm{end}}}^{\phi}
\left|\frac{V_\mathrm{E}}{V'_\mathrm{E}}\right|Q\,d\tilde\phi=
\frac{\kappa}{\sqrt{2}}\int\limits_{\phi_{\mathrm{end}}}^{\phi}
\left(\frac{d\varphi}{d\tilde\phi}\right)\frac{d\tilde\phi}{\sqrt{\epsilon(\tilde\phi)}}\,,
\label{Ne}
\end{equation}
where $\phi_{\mathrm{end}}$ is the value of the field at the end of inflation, defined by $\epsilon=1$. The number of e-foldings must be matched with the appropriate normalization of the data set
and the cosmic history, a typical value being $50\leqslant N_e\leqslant65$.

The tensor-to-scalar ratio $r$, the scalar spectral index of the
primordial curvature fluctuations~$n_\mathrm{s}$, and the
associated running of the spectral index $\alpha_\mathrm{s}$, are
given, to very good approximation, by
\begin{equation}\label{ns}
r = 16 \epsilon\,,\qquad n_\mathrm{s} - 1 \simeq - 6 \epsilon + 2
\eta\, , \qquad
\alpha_\mathrm{s} \equiv \frac{d n_\mathrm{s}}{d \ln k}
\simeq 16\epsilon \eta - 24 \epsilon^2 - 2 \zeta^2
\, .
\end{equation}

Planck2013 temperature anisotropy measurements~\cite{Planck2013} combined with the WMAP large-angle polarization, constrain the scalar spectral index to $n_\mathrm{s}=0.9603\pm0.0073$.
Our goal is to check the possibility to get a value of $n_s$ in the models investigated here.
We describe the inflationary dynamics for two considered models that have
unstable de Sitter solutions with $U_f>0$. Observe that the existence of an unstable de Sitter solution may not be a necessary condition for inflation. Stable de Sitter solutions may be the basis for eternal inflation. From another side, stable de Sitter solutions may appear to be
true inflationary solutions subject to the condition that the exit from inflation
occurs due to some other scenario.  We do not consider such inflationary
scenarios, that can be suitable for the $SU(2)$ model, in this paper.

Note that for the $SU(5)$ model, the de Sitter solutions obtained exist for $U_f<0$, as well. In this case one cannot use the conformal transformation to formulate the model in the Einstein frame. Also, stability conditions can be violated~\cite{Polarski}. The possibility
 to develop the inflationary scenario without pathologies in this case demands a more detailed analysis, which will be carried out in future works.

\subsection{Scalar electrodynamics}
\begin{figure}[!ht]
\centering
\includegraphics[width=72mm]{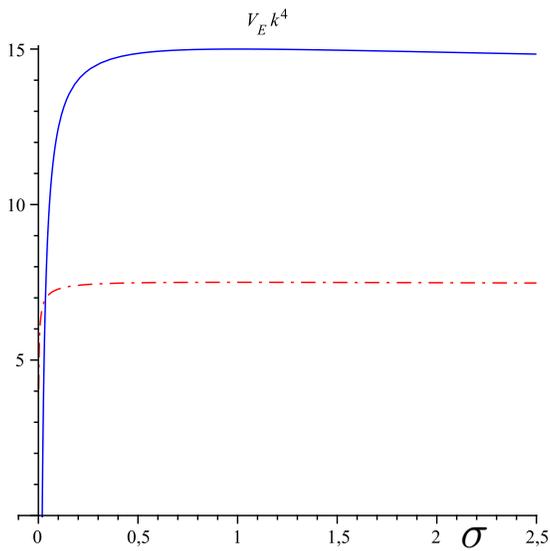}
\caption{The potential $V_\mathrm{E}(\sigma)$ multiplied by $\kappa^4$ in the scalar electrodynamics model. In both pictures $\lambda=18 e^2\xi$. This means that $\sigma_f=1$. The parameter values are $e=10$, $\xi=10$ (red dashed  line) and $e=10$, $\xi=5$ (blue solid line). Note that the potential $V_\mathrm{E}$ with such choice of $\lambda$ depends on the combination $\xi/e^2$ only.}
\label{VESCALED}
\end{figure}

For scalar electrodynamics, the functions $V$ and $U$ are given by (\ref{VUSCED}); therefore,
\begin{equation}\label{VESCED}
V_\mathrm{E}=\frac{ 16\pi^2\left(8\pi^2\lambda+9e^4\ln(\sigma^2)\right)}{3\kappa^4\left(32\xi\pi^2+e^2\ln(\sigma^2)\right)^2}\,,
\end{equation}
\begin{equation}\label{QSCED}
    Q=\frac{6e^4\ln\left(\sigma^2\right)^2+4[3e^2+8\pi^2(1+12\xi)]e^2
    \ln(\sigma^2)+6e^4+384\pi^2e^2\xi+1024\pi^4\xi(1+6\xi)}{
    \kappa^2\mu^2\sigma^2[32\pi^2\xi+e^2\ln(\sigma^2)]^2}\,,
\end{equation}
where we use the dimensionless quantity $\sigma\equiv\phi/\mu$. Note that the slow-roll parameters
$\epsilon$ and $\eta$ do not depend on the dimensionless combination $\kappa\mu$.
We choose the parameters so that $U_f>0$, what means $\lambda<36 e^2\xi$. Note that the de Sitter solutions correspond to the condition $V'_\mathrm{E}(\phi_f)=0$. Solutions are unstable; so, in this point
the potential $V_\mathrm{E}$ has a maximum. In Fig.~\ref{VESCALED} we see that the potential is very flat near the maximum and decreases more rapidly than $\sigma$, when it tends to zero.

In our calculations, we put $\lambda=18 e^2\xi$ what gives $\phi_f=\pm\mu$ ($\sigma_f=\pm 1$). At the de Sitter point $\phi_f$ the value of the RG-improved potential coincides with the value of the tree-level one. We consider positive values for $\sigma$ and for the parameters in the action.
The corresponding inflationary parameters are listed in Table~\ref{ScalarElectroDinPar}.
\begin{table}[h]
\begin{center}
\caption{Parameter values for the scalar electrodynamics inflationary scenario.}
\begin{tabular}{|c|c|c|c|c|c|c|c|}
  \hline
   $\xi$& $e$&  $\sigma_{end}$ ($\epsilon=1$)& $N_e$ & $\sigma_N$ & $n_\mathrm{s}$ & $r$ & $\alpha_\mathrm{s}$ \\
   \hline
  $5$ & $10$ & $0.02415237908$ & $50$ & $0.20044$ & $0.962$ & $0.013$ & $ 0.0013 $ \\
  $5$ & $10$ & $0.02415237908$ & $55$ & $0.21544$ & $0.965$ & $0.011$ & $ 0.0011 $ \\
  $5$ & $10$ & $0.02415237908$ & $60$ & $0.23023$ & $0.967$ & $0.009$ & $  0.0009 $ \\
  $5$ & $10$ & $0.02415237908$ & $65$ & $0.24485$ & $0.969$ & $0.008$ & $  0.0008 $ \\
  \hline
  $10$ & $10$ & $0.0004788703192$ & $50$ & $0.007189$ & $0.967$ & $0.028$ & $ 0.0036$ \\
  $10$ & $10$ & $0.0004788703192$ & $55$ & $0.007974$ & $0.965$ & $0.024$ & $ 0.0032 $ \\
  $10$ & $10$ & $0.0004788703192$ & $60$ & $0.008848$ & $0.967$ & $0.022$ & $ 0.0026$ \\
  $10$ & $10$ & $0.0004788703192$ & $65$ & $0.009756$ & $0.974$ & $0.019$ & $  0.0025$ \\
  \hline
\end{tabular}
\label{ScalarElectroDinPar}
\end{center}
\end{table}

One can see that, for the values of the parameters  $e$ and  $\xi$ presented in Table~\ref{ScalarElectroDinPar}, the corresponding values of $n_s$ and $r$ are in good agreement with the observational data~\cite{Planck2013,Seljak}. The potential $V_\mathrm{E}(\sigma)$ with these values of parameters are presented in Fig.~\ref{VESCALED}.

\subsection{The $SU(5)$ model}
In the case of a $SU(5)$ RG-improved potential,
\begin{equation}
\label{Vein}
V_\mathrm{E}=
\frac{135g^2(\breve{\Theta}-1)\breve{\Theta}^{5/4}}{32\kappa^4
\left(\breve{\Theta}^{9/8}+6\xi-1\right)^2}\,,
\end{equation}
\begin{equation}
Q=\frac{ 4(  \breve{\Theta}^{9/8}+6\xi-1)+{\frac {15}{128{\pi }^{2}}}\, \left( 15g^2\breve{\Theta}^{1/8}+16{\pi }^{2}\, \left( \breve{\Theta}^{9/8}
+6\,\xi-1\right)\right) ^{2
}  }{ 5\left(\breve{\Theta}^{9/8}+6\,\xi-1 \right)^{2}{\kappa}^{2}{\phi}^{2}}\,.
\end{equation}

For the $SU(5)$ RG-improved potential (\ref{SU5pot}), the function $\varphi(\phi)$
cannot be written in closed form.
Because of this reason, we write the slow-roll  parameters as functions of the
Jordan-frame scalar field $\phi$. The slow-roll parameters read as follows:
\begin{equation*}
\epsilon=\frac{125{g}^{4}\left( 4\,{\breve{\Theta}}^{9/8}-5(6\xi-1)+9\breve{\Theta}(6\xi-1)\right)^{2}}{288\pi^4
\left(\breve{\Theta}-1 \right)^{2}\breve{\Theta}^2
 \left[4\left( {\breve{\Theta}}^{9/8}+6\,\xi-1 \right) +{\frac {15}{128\pi^2}}\, \left( 15\,{g}^{2}{\breve{\Theta}}^{9/8}+16\pi^2\left({\breve{\Theta}}^{9/8}+6\xi-1\right)^{2}\right) \right] },
\end{equation*}
and
\begin{equation}
N_e={\frac {36}{125}}\int\limits^{\breve{\Theta}_N}_{\breve{\Theta}_{end}}\frac{{\pi }^{4} \left( \breve{\Theta}-1 \right) \breve{\Theta}\, \left(
4\left({\breve{\Theta}}^{{\frac {9}{8}}}+6\xi-1\right)+{\frac {15}{128}}\, \left(
{\frac {15\,{\breve{\Theta}}^{\frac18}{g}^{2}}{{\pi }^{2}}}+16({\breve{\Theta}}^{{\frac {9
}{8}}}+6\xi-1) \right) ^{2} \right)}{ {g}^{4}\left( {\breve{\Theta}
}^{{\frac {9}{8}}}+6\,\xi-1 \right)\left( (9\,\breve{\Theta}-5)\left( 6
\,\xi-1 \right) +4\,{\breve{\Theta}}^{{\frac {9}{8}}}\right)}d\breve{\Theta}\,.
\end{equation}
We see that the slow-roll parameters and $N_e$ depend on the dimensionless function $\breve{\Theta}$.
\begin{figure}[!ht]
\centering
\includegraphics[width=72mm]{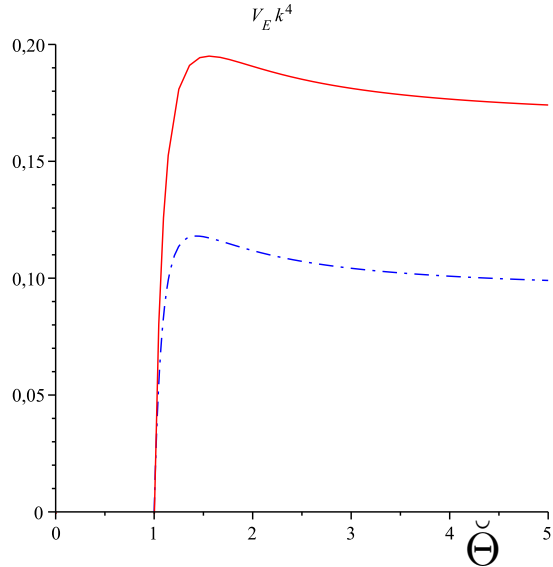}
\caption{The potential $V_\mathrm{E}(\breve{\Theta})$ multiplied by $\kappa^4$ in the $SU(5)$ model at $\xi=0.04$, $g=0.15$ (blue dashed line) and at $\xi=0.045$, $g=0.2$ (red solid line).}
\label{SU(5)Veff}
\end{figure}

The potential $V_\mathrm{E}$ for  $\xi=0.04$ and $\xi=0.045$ is plotted in Fig.~\ref{SU(5)Veff}.
The corresponding inflationary parameters are listed in Table~\ref{SU5par}.
\begin{table}[h]
\begin{center}
\caption{Parameter values for the $SU(5)$ inflationary scenario.}
\begin{tabular}{|c|c|c|c|c|c|c|c|}
  \hline
  $\xi$ & $g$ & $N$ & $\breve{\Theta}_{end}(\epsilon=1)$ & $\breve{\Theta}_N$ & $n_s$ & $r$& $\alpha$  \\
  \hline
  $0.04$ & $0.15$ & $50$ & $1.000868906$ & $1.0121 $ & $0.963 $ & $0.070$ & $0.00731$  \\
  $0.04$ & $0.15$ & $55$ & $1.000868906$ & $1.0126$ & $0.965$ & $0.063$ & $0.00643$  \\
  $0.04$ & $0.15$& $60$ & $1.000868906$ & $1.0132 $ & $0.968 $ & $0.058$ & $0.00660$   \\
  $0.04$ & $0.15$& $65$ & $1.000868906$ & $1.0137$ & $0.969 $ & $0.0535$ & $0.00540$  \\
  \hline
 $0.045$ & $0.2$ & $50$ & $1.001564816$ & $1.02152$ & $0.958$ & $0.066$ & $ 0.00699$  \\
 $0.045$ & $0.2$ & $55$ & $1.001564816$ & $1.02252$ & $0.960$ & $0.0595$ & $ 0.00638$   \\
 $0.045$ & $0.2$ & $60$ & $1.001564816$ & $1.023475$ & $0.963$ & $0.054$ & $ 0.00579$  \\
 $0.045$ & $0.2$ & $65$ & $1.001564816$ & $1.024388$ & $0.965$ & $0.0495$ & $ 0.00548$  \\
  \hline
\end{tabular}
\label{SU5par}
\end{center}
\end{table}

 The resulting joint BICEP2+Planck2013 analysis yields that the upper limit of the tensor-to-scalar ratio is $r<0.11$, a slight improvement relative to the Planck analysis alone, which gives $r<0.13$ (95\% c.l.)~\cite{Seljak}. We do see that the inflationary parameters of the model considered are in very good agreement with the observational data.

\section{Conclusions}

In this paper we have considered the possibility to construct inflationary models for the renormali\-zation-group improved potentials corresponding to scalar electrodynamics and to the $SU(2)$ and $SU(5)$ models. In all cases, the tree-level potential is $\lambda \phi^4- \xi\phi^2R$, what corresponds to the cosmological constant in the Einstein frame, and is in no case suitable for inflation.
The standard way to get an inflationary model is to add the Hilbert--Einstein term to the action~\cite{HiggsInflation,DeSimone:2008ei}. Actually we did not add this term here, but included instead the quantum corrections to the potential coming from to the RG-equation. We then analyzed the corresponding inflationary scenario with unstable de Sitter solutions, only. This means that the corresponding potential in the Einstein frame should have a maximum. We have found that, for some reasonable values of the parameters, this is indeed the case, both for scalar electrodynamics and for the $SU(5)$ model, and that the corresponding values of the coupling function are indeed positive, $U_f>0$. For the finite $SU(2)$ model de Sitter solutions exist, if it corresponds to the model with a minimally coupled scalar field and a cosmological constant; thus, this case is not suitable for inflation. In the $SU(2)$ gauge model there exist stable de Sitter solutions for $U_f>0$, only. Note also that a stable de Sitter solution may appear to be a true inflationary solution, but only under the condition that the exit from inflation occurs according to some other particular scenarios; but we did not consider such inflationary scenarios in this paper.

In the inflationary models, both for scalar electrodynamics and the $SU(5)$ RG-improved potentials,  we have got that these models are in good agreement with the most recent observational data~\cite{Planck2013,Seljak} provided some reasonable values are taken for the parameters.

Our study indicates that inflation could well be caused by quantum effects of the
 scalar sector of some convenient GUT theory. We believe this is quite a remarkable result. In this respect, it would be of interest to investigate the possibility of inflation in GUTs with other gauge
 groups, as the exceptional $E8$ group, or GUTs which proceed from the string
 framework. From another side, adding a RG-improved effective potential to the
 classical GR action may lead to a qualitative change of the inflationary
 dynamics which occur in such models. This issue will be discussed elsewhere.

\medskip

\noindent {\bf Acknowledgements}. E.E. and S.D.O. are supported in part by MINECO (Spain), project FIS2010-15640, and by the CPAN Consolider Ingenio Project. E.O.P. and S.Yu.V. are partially supported by RFBR grant 14-01-00707 and by the Russian Ministry of Education and Science under grant NSh-3042.2014.2.


\begin{thebibliography}{72}
\bibitem{WMAP}
 D.N.~Spergel {\it et al.}  [WMAP Collaboration],
 Astrophys.\ J.\ Suppl. {\bf 148} (2003)  175--194
 (arXiv:astro-ph/0302209);  \\
 D.N.~Spergel {\it et al.}  [WMAP Collaboration],
 Astrophys.\ J.\ Suppl. {\bf 170} (2007)
 377 (arXiv:astro-ph/0603449); \\
E.~Komatsu {\it et al.}  [WMAP Collaboration],
  Astrophys.\ J.\ Suppl.  {\bf 192} (2011) 18
  (arXiv:1001.4538);\\
  G.~Hinshaw {\it et al.}  [WMAP Collaboration],
Astrophys.\ J.\
Suppl.\ {\bf 208} (2013) 19 (arXiv:1212.5226)

\bibitem{Planck2013}
P.A.R. Ade, {\it et. al.} [Planck Collaboration],
arXiv:1303.5076;\\
P.A.R. Ade, {\it et. al.} [Planck Collaboration],
arXiv:1303.5082;\\
P.A.R. Ade, {\it et. al.} [Planck Collaboration],
arXiv:1303.5084

\bibitem{BICEP2}
 P.A.R.~Ade {\it et al.}  [BICEP2 Collaboration],
  arXiv:1403.3985


\bibitem{Seljak}
  M.J.~Mortonson and U.~Seljak,
  arXiv:1405.5857

\bibitem{Starobinsky:1979ty}
 A.A.~Starobinsky,
JETP Lett.\ {\bf 30} (1979) 682 [Pisma Zh.\ Eksp.\
Teor.\ Fiz.\ {\bf 30} (1979) 719--723];\\
  A.A.~Starobinsky,
    Phys.\ Lett. B {\bf 91} (1980)  99--102;\\
A.A.~Starobinsky, Lect. Notes in Phys. \textbf{246}
(1986) 107

\bibitem{Mukhanov:1981xt}
  V.F.~Mukhanov and G.V.~Chibisov,
JETP Lett.\ {\bf 33} (1981) 532--535, [Pisma Zh.\
Eksp.\ Teor.\ Fiz.\ {\bf 33} (1981) 549--553].

  \bibitem{Guth:1980zm}
  A.H.~Guth,
     Phys. Rev. D {\bf 23} (1981) 347

  \bibitem{Linde:1981mu}
  A.D.~Linde,
Phys.\ Lett.\ B {\bf 108} (1982) 389;\\
A.D.~Linde,
Phys.\ Lett.\ B {\bf 129} (1983) 177



\bibitem{SU5inflation}
 K. Olive, Phys. Rep. {\bf 190} (1990) 309;\\
E.W. Kolb and M.S. Turner, {\it The Early Universe},
Addison-Wesley, 1990; E.W. Kolb, Phys. Scr. {\bf T36} (1991)
199.

\bibitem{Albrecht:1982wi}
A.~Albrecht and P.J.~Steinhardt,
Phys. Rev. Lett.  {\bf 48} (1982)  1220

\bibitem{inflation2}
J.E. Lidsey, A.R. Liddle, E.W. Kolb, E.J. Copeland, T. Barreiro,
and M. Abney,
Rev. Mod. Phys. \textbf{69} (1997) 373--410
(arXiv:astro-ph/9508078);\\
  B.~A.~Bassett, S.~Tsujikawa and D.~Wands,
  Rev.\ Mod.\ Phys.\  {\bf 78} (2006) 537 (astro-ph/0507632);\\
C.M. Peterson, M. Tegmark,
Phys. Rev. D \textbf{83} (2011) 023522
(arXiv:1005.4056);\\
Shi Pi, M. Sasaki,
J. Cosmol. Astropart. Phys. \textbf{1210} (2012) 051
(arXiv:1205.0161);\\
P.~Creminelli, D.L.~Nacir, M.~Simonovic, G.~Trevisan and
M.~Zaldarriaga,
  arXiv:1405.6264

\bibitem{nonmin-infl}
B.L. Spokoiny,
Phys. Lett. B \textbf{147} (1984) 39--43;\\
T. Futamase and K.-i.  Maeda,
Phys. Rev. D \textbf{39} (1989) 399--404;\\
R. Fakir and W.G. Unruh,
Phys. Rev. D \textbf{41} (1990) 1783--1791;\\
M.V.~Libanov, V.A.~Rubakov and P.G.~Tinyakov,
Phys. Lett. B {\bf 442} (1998) 63 (arXiv:hep-ph/9807553);\\
 R.~Kallosh, A.~Linde and D.~Roest,
  arXiv:1407.4471
\bibitem{Salopek}
D.S. Salopek, J.R. Bond and J.M. Bardeen,
Phys. Rev. D \textbf{40} (1989) 1753--1788;\\
D.S. Salopek, J.R. Bond,
Phys. Rev.  D {\bf 42} (1990) 3936

\bibitem{OdintsovNOnmin}
T. Muta, S.D. Odintsov,
Mod. Phys. Lett. A \textbf{6} (1991) 3641--3646;\\
S. Mukaigawa, T. Muta, S.D. Odintsov,
Int. J. Mod. Phys. A \textbf{13} (1998) 2739--2746 (arXiv:hep-ph/9709299)

\bibitem{nonmin-quant}
A.O. Barvinsky  and A.Yu. Kamenshchik,
Phys. Lett.  B {\bf 332} (1994) 270  (arXiv:gr-qc/9404062)
\bibitem{Cerioni}
A. Cerioni, F. Finelli, A. Tronconi and G. Venturi,
Phys. Lett. B \textbf{681} (2009) 383--386
(arXiv:0906.1902); \\
A. Cerioni, F. Finelli, A. Tronconi and G. Venturi,
Phys. Rev. D \textbf{81} (2010) 123505 (arXiv:1005.0935);
\\
A. Tronconi and G. Venturi,
Phys. Rev. D \textbf{84} (2011) 063517 (arXiv:1011.39580)

\bibitem{HiggsInflation}
F.L.~Bezrukov and M.~Shaposhnikov,
Phys. Lett. B \textbf{659} (2008) 703 (arXiv:0710.3755);
\\
A.O.~Barvinsky, A.Y.~Kamenshchik, and A.A.~Starobinsky,
J. Cosmol. Astropart. Phys. {\bf 0811} (2008) 021
(arXiv:0809.2104);\\
 F.~Bezrukov, D.~Gorbunov and M.~Shaposhnikov,
J. Cosmol. Astropart. Phys. {\bf 0906} (2009) 029,
arXiv:0812.3622;\\
F.L.~Bezrukov, A.~Magnin, and M.~Shaposhnikov,
Phys. Lett. B {\bf 675} (2009) 88 (arXiv:0812.4950);\\
 A.O.~Barvinsky, A.Y.~Kamenshchik, C.~Kiefer, A.A.~Starobinsky,
and C.F.~Steinwachs,
J. Cosmol. Astropart. Phys. \textbf{0912} (2009) 003
(arXiv:0904.1698);\\
J.~Garcia-Bellido, D.G.~Figueroa, and J.~Rubio,
Phys. Rev. D \textbf{79} (2009) 063531
(arXiv:0812.4624);\\
R.N. Lerner and J. McDonald,
J. Cosmol. Astropart. Phys. \textbf{1004} (2010) 015
(arXiv:0912.5463);\\
F.L.~Bezrukov, A.~Magnin, M.~Shaposhnikov and S.~Sibiryakov,
J. High Energy Phys. {\bf 1101} (2011) 016
(arXiv:1008.5157)\\
A.O.~Barvinsky, A.Yu.~Kamenshchik, C.~Kiefer, A.A.~Starobinsky,
and C.F.~Steinwachs,
Eur. Phys. J. C \textbf{72} (2012) 2219
(arXiv:0910.1041);\\
F.L.~Bezrukov and D.S.~Gorbunov,
  Phys.\ Lett.\ B {\bf 713} (2012) 365 (arXiv:1111.4397);\\
F. Bezrukov,
Class. Quant. Grav. {\bf 30} (2013) 214001
(arXiv:1307.0708);\\
K.~Allison,
  J. High Energy Phys. {\bf 1402} (2014) 040 (arXiv:1306.6931);\\
Y.~Hamada, H.~Kawai, K.-y.~Oda and S.C.~Park,
  Phys. Rev. Lett. \textbf{112} (2014) 241301 (arXiv:1403.5043);\\
   J.~Ren, Z.-Z.~Xianyu, H.-J.~He,
J. Cosmol. Astropart. Phys. {\bf 1406} (2014) 032 (arXiv:1404.4627);\\
  Hong-Jian He and Zhong-Zhi Xianyu,
  arXiv:1405.7331 [hep-ph].

\bibitem{DeSimone:2008ei}
  A.~De Simone, M.P.~Hertzberg and F.~Wilczek,
  Phys.\ Lett.\ B {\bf 678} (2009) 1
  (arXiv:0812.4946)

\bibitem{GB2013}
F.~Bezrukov, D.~Gorbunov,
J. High Energy Phys.
\textbf{1307} (2013) 140  (arXiv:1303.4395)

\bibitem{KL2013}
R.~Kallosh, A.~Linde,
J. Cosmol. Astropart. Phys. {\bf 1306} (2013) 027
(arXiv:1306.3211)

\bibitem{Lindebook}
A.D.~Linde, \emph{Particle Physics and Inflationary Cosmology},
Contemporary Concepts in Physics \textbf{5} (1990) 1--362, Harwood Academic, New York,
(arXiv:hep-th/0503203)

\bibitem{19} E.W. Kolb and M.S. Turner, {\it The Early
Universe}, Addison-Wesley, Reading, MA, 1990.

\bibitem{Inflation_review}
 A.D.~Linde,
  Lect.\ Notes Phys.\  {\bf 738} (2008) 1
  (arXiv:0705.0164);\\
   J.~Martin, C.~Ringeval, V.~Vennin,
  Phys.\ Dark Univ.\  (2014)  [arXiv:1303.3787];\\
 A.D.~Linde,
  arXiv:1402.0526


\bibitem{Wan:2014fra}
  Y.~Wan, S.~Li, M.~Li, T.~Qiu, Y.~Cai and X.~Zhang,
  arXiv:1405.2784;\\
 C. Germani, Y. Watanabe and N. Wintergerst,
arXiv:1403.5766

\bibitem{Hossain:2014ova}
  M.~Wail Hossain, R.~Myrzakulov, M.~Sami and E.~N.~Saridakis,
  arXiv:1405.7491

\bibitem{Garcia-Bellido:2014eva}
  J.~Garcia-Bellido, D.~Roest, M.~Scalisi and I.~Zavala,
  arXiv:1405.7399

\bibitem{Barranco:2014ira}
  L.~Barranco, L.~Boubekeur and O.~Mena,
  arXiv:1405.7188

\bibitem{Martin:2014lra}
  J.~Martin, C.~Ringeval, R.~Trotta and V.~Vennin,
  arXiv:1405.7272

\bibitem{Gao:2014pca}
  Q.~Gao, Y.~Gong and T.~Li,
  arXiv:1405.6451

\bibitem{CH-CHZ}
  C.~Cheng and Q.~-G.~Huang,
  arXiv:1404.1230;\\
  C.~Cheng, Q.~-G.~Huang and W.~Zhao,
  arXiv:1404.3467

\bibitem{Hamada:2014xka}
  Y.~Hamada, H.~Kawai and K.-y.~Oda,
  arXiv:1404.6141

\bibitem{KS-KSY-HKSY}
  T.~Kobayashi and O.~Seto,
  Phys.\ Rev.\ D {\bf 89} (2014) 103524
  (arXiv:1403.5055);\\
  T.~Kobayashi and O.~Seto,
  arXiv:1404.3102;\\
 T.~Higaki, T.~Kobayashi, O.~Seto and Y.~Yamaguchi,
  arXiv:1405.0775

\bibitem{Inagaki:2014wva}
  T.~Inagaki, R.~Nakanishi and S.D.~Odintsov,
  arXiv:1408.1270


\bibitem{Bamba:2014daa}
  K.~Bamba, Sh.~Nojiri and S.D.~Odintsov,
  arXiv:1406.2417

\bibitem{Cervantes-Cota1995}
   J.L.~Cervantes-Cota and H.~Dehnen,
Nucl.\ Phys. B {\bf 442} (1995) 391
(arXiv:astro-ph/9505069)

\bibitem{Lyth:1998xn}
  D.H.~Lyth and A.~Riotto,
  Phys.\ Rept.\  {\bf 314} (1999) 1--146 (arXiv:hep-ph/9807278)



\bibitem{SUSEinflation}
B.A.~Ovrut and P.J.~Steinhardt,
  Phys.\ Lett.\ B {\bf 147} (1984) 263;\\
G.R.~Dvali,
  Phys.\ Lett.\ B {\bf 387} (1996) 471
  (arXiv:hep-ph/9605445);\\
   L.~Alvarez-Gaume, C.~Gomez and R.~Jimenez,
  J. Cosmol. Astropart. Phys. {\bf 1103} (2011) 027
  (arXiv:1101.4948);\\
   C.~Pallis,
  arXiv:1403.5486 [hep-ph]

\bibitem{GUT_Inflation}
 J.L.~Cervantes-Cota and H.~Dehnen,
Phys.\ Rev. D {\bf 51} (1995) 395
(arXiv:astro-ph/9412032);\\
  S.~Dimopoulos, G.R.~Dvali and R.~Rattazzi,
  Phys.\ Lett.\ B {\bf 410} (1997) 119
  (arXiv:hep-ph/9705348);\\
 M.B.~Einhorn and D.R.T.~Jones,
  J. Cosmol. Astropart. Phys. {\bf 1211} (2012) 049
  (arXiv:1207.1710);\\
  F.~Brummer, V.~Domcke and V.~Sanz,
  arXiv:1405.4868

\bibitem{BOS} I.L. Buchbinder, S.D. Odintsov and I.L. Shapiro,
 \textit{Effective Action in Quantum Gravity}, IOP Publishing, Bristol
and
Philadelphia, 1992.\\
I.L. Buchbinder,
 S.D. Odintsov,
 I.L. Shapiro,
\textit{Renormalization group approach to quantum field theory in curved space-time},
Rivista Nuovo Cimento Ser. 3, V.~\textbf{12}, Iss. 10 (1989) 1--112;\\
S.D. Odintsov,
Fortsch. Phys. \textbf{39} (1991) 621--641

\bibitem{BO1985}
 I.L. Buchbinder, S.D. Odintsov,
  Class. Quant. Grav. \textbf{2} (1985) 721--731

\bibitem{Elizalde:1993ee}
  E.~Elizalde and S.D.~Odintsov,
Phys.\ Lett.\ B {\bf 303} (1993) 240  (arXiv:hep-th/9302074);\\
  E. Elizalde and S.D. Odintsov,
 Phys. Lett. B \textbf{321} (1994) 199--204 (arXiv:hep-th/9311087)


\bibitem{Freedman}
D.Z. Freedman, I.J. Muzinich and E.J. Weinberg, Ann. Phys. 87 (1974) 95;\\
D.Z. Freedman and E.J. Weinberg, Ann. Phys. 87 (1974) 354;\\
D.Z. Freedman and S.Y. Pi, Ann. Phys. 91 (1975) 442

\bibitem{BirrellDavies} N.D. Birrell and P.C.W. Davies,
\textit{Quantum Fields in Curved Space}, Cambridge University Press, Cambridge,
England, 1982

\bibitem{Elizalde:1994im}
  E.~Elizalde and S.D.~Odintsov,
  Phys.\ Lett.\ B {\bf 333} (1994) 331
  hep-th/9403132.

\bibitem{Cooper:1982du}
  F.~Cooper and G.~Venturi,
 Phys. Rev.  D \textbf{24} (1981) 3338


\bibitem{Kaiser}
D.I.~Kaiser,
Phys. Lett. B \textbf{340} (1994) 23--28
(arXiv:astro-ph/9405029);\\
D.I.~Kaiser,
Phys. Rev. D {\bf 52} (1995) 4295
(arXiv:astro-ph/9408044)

\bibitem{KKhT} A.Yu. Kamenshchik, I.M. Khalatnikov, and A.V.
Toporensky,
Int. J. Mod. Phys. D \textbf{6} (1997) 649--672
(arXiv:gr-qc/9801039)

\bibitem{Polarski}
 R.~Gannouji, D.~Polarski, A.~Ranquet, and A.A.~Starobinsky,
J. Cosmol. Astropart. Phys. {\bf 0609} (2006) 016
(arXiv:astro-ph/0606287)

\bibitem{Elizalde}
E. Elizalde, Sh. Nojiri, and S.D. Odintsov,
Phys. Rev. D \textbf{70} (2004) 043539
(arXiv:hep-th/0405034);\\
E.~Elizalde, Sh.~Nojiri, S.D.~Odintsov, D.~Saez-Gomez and
V.~Faraoni,
Phys. Rev. D \textbf{77} (2008) 106005
(arXiv:0803.1311);\\
D.~Saez-Gomez,
Phys. Rev. D {\bf 85} (2012) 023009 (arXiv:1110.6033)


\bibitem{CervantesCota:2010cb}
J.L.~Cervantes-Cota, R.~de Putter, and E.V.~Linder,
J. Cosmol. Astropart. Phys. {\bf 1012} (2010) 019
(arXiv:1010.2237)

 \bibitem{Kamenshchik:2012rs}
  A.Y.~Kamenshchik, A.~Tronconi and G.~Venturi,
Phys. Lett. B {\bf 713} (2012) 358 (arXiv:1204.2625)

\bibitem{KTV2011}
A.Yu. Kamenshchik, A. Tronconi, and G. Venturi,
Phys. Lett. B \textbf{702} (2011) 191--196
(arXiv:1104.2125);\\
A.Yu.~Kamenshchik, A.~Tronconi, G.~Venturi, and S.Yu.~Vernov,
Phys. Rev. D \textbf{87} (2013) 063503 (arXiv:1211.6272)

\bibitem{Sami:2012uh}
  M.~Sami, M.~Shahalam, M.~Skugoreva, and A.~Toporensky,
Phys. Rev. D \textbf{86} (2012) 103532
(arXiv:1207.6691);\\
A.Yu.~Kamenshchik, E.O.~Pozdeeva, A.~Tronconi, G.~Venturi and
S.Yu.~Vernov,
Class. Quant. Grav. \textbf{31} (2014) 105003
(arXiv:1312.3540);\\
 M.A.~Skugoreva, A.V.~Toporensky, and S.Yu.~Vernov,
Phys. Rev. D \textbf{90} (2014) to be published,  arXiv:1404.6226

\bibitem{ABGV}
I.Ya. Aref'eva, N.V. Bulatov, R.V. Gorbachev, S.Yu. Vernov,
Class. Quant. Grav. \textbf{31} (2014) 065007
(arXiv:1206.2801)




\bibitem{Book-Capozziello-Faraoni}
S.~Capozziello and V.~Faraoni, \emph{Beyond Einstein Gravity: A
Survey of Gravitational Theories for Cosmology and
Astrophysics}, Fund. Theor. Phys. \textbf{170}, Springer, New
York, 2011;\\
S. Capozziello and M. De Laurentis,
Phys. Rep. {\bf 509} (2011) 167--321 (arXiv:1108.6266)

\bibitem{Fujii_Maeda} Y. Fujii and K. Maeda, \emph{The
Scalar--Tensor Theory of Gravitation}, Cambridge University
Press,
Cambridge, 2004

\bibitem{NO-rev}
 S.~Nojiri and S.D.~Odintsov,
Int.\ J.\ Geom.\ Meth.\ Mod.\ Phys. {\bf 4} (2007)
115--146 (arXiv:hep-th/0601213);\\
S.~Nojiri and S.D.~Odintsov,
Phys. Rept. \textbf{505} (2011) 59--144 (arXiv:1011.0544)

\bibitem{Lyapunov} A.M. Lyapunov, \emph{Stability of motion},
Academic
Press, New-York and London, 1966 (in English); A.M.  Lyapunov,
\emph{General problem of stability of motion}, GITTL,
Moscow--Leningrad,  1950 (in Russian)

\bibitem{Pontryagin} L.S. Pontryagin, \emph{Ordinary
Differential Equations},
Adiwes International Series in Mathematics. Addison-Wesley Publ.
Comp.,
London--Paris, 1962 (in English), ''Nauka'', Moscow, 1982 (in
Russian)

\bibitem{PV2014}
 E.O.~Pozdeeva and S.Yu.~Vernov,
AIP Conf. Proc. \textbf{1606} (2014) 48--58 (arXiv:1401.7550)

\bibitem{Coleman} S. Coleman and E. Weinberg, Phys. Rev. D {\bf
7} (1973) 1888.

\bibitem{Sher} M. Sher, Phys. Rep. {\bf 179} (1989) 273;\\
M. Sher, Phys. Lett. B {\bf 135} (1984) 52.

\bibitem{SU2} B.L. Voronov and I.V. Tyutin, Yad. Fiz. (Sov. J.
Nucl. Phys.) {\bf 23} (1976) 664;\\
I.L. Buchbinder and S.D. Odintsov,
Yad. Fiz. (Sov. J. Nucl. Phys.) {\bf 40} (1984) 1338.

\bibitem{Odintsov7}  S.D. Odintsov and I.L. Shapiro, Mod. Phys. Lett.
{\bf A4} (1989) 1479;\\
 S.D. Odintsov, D.J. Toms, I.L. Shapiro,
 Int. J. Mod. Phys. A \textbf{6} (1991) 1829--1834

\bibitem{Ermushev8} A.V. Ermushev, D.I. Kazakov and O.V.
Tarasov,
Nucl. Phys. B {\bf 281} (1987) 72;\\ D. Kapetanakis, M.
Mondrag\'on
and G. Zoupanos, Z. Phys. C {\bf 60} (1993) 181.

\bibitem{9} I.L. Buchbinder, S.D. Odintsov and I.M. Lichtzier,
Class. Quant. Grav. {\bf 6} (1989) 6055.

\bibitem{4} M. B\"{o}hm and A. Denner, Nucl. Phys. B {\bf 282}
(1987) 206.

\bibitem{8} H. Georgi and S.L. Glashow, Phys. Rev. Lett. {\bf
32}
(1974) 438.

\bibitem{Liddle:1994dx}
 A.R.~Liddle, P.~Parsons and J.D.~Barrow,
  Phys.\ Rev.\ D {\bf 50} (1994) 7222
 (arXiv:astro-ph/9408015)


\end{thebibliography}
\end{document}